\documentclass[aps,prl,nofootinbib,superscriptaddress,twocolumn,amsmath,amssymb]{revtex4-1}
\usepackage{amsmath,amsfonts,amssymb,graphicx}
\usepackage[usenames]{color}
\usepackage{epsfig}
\usepackage{epstopdf}
\newcommand{\be}{\begin{equation}}
\newcommand{\ee}{\end{equation}}
\newcommand{\bey}{\begin{eqnarray}}
\newcommand{\eey}{\end{eqnarray}}
\newcommand{\bw}{\begin{widetext}}
\newcommand{\ew}{\end{widetext}}

\newcommand{\ba}{\begin{array}}
\newcommand{\ea}{\end{array}}
\newcommand{\bi}{\begin{itemize}}
\newcommand{\ei}{\end{itemize}}
\newcommand{\bem}{\begin{enumerate}}
\newcommand{\eem}{\end{enumerate}}

\newcommand{\hefei}{Department of Modern Physics, University of Science and Technology of China, Hefei 230026, China}
\newcommand{\como}{Center for Nonlinear and Complex Systems, Dipartimento di Scienza e Alta Tecnologia,
Universit\`a degli Studi dell'Insubria, via Valleggio 11, 22100 Como, Italy}
\newcommand{\infn}{Istituto Nazionale di Fisica Nucleare, Sezione di Milano, via Celoria 16, 20133 Milano, Italy}
\newcommand{\brazil}{International Institute of Physics, Federal University of Rio Grande do Norte,
Campus Universit\'ario - Lagoa Nova, CP. 1613, Natal, Rio Grande Do Norte 59078-970, Brazil}
\newcommand{\NEST}{NEST, Istituto Nanoscienze-CNR, I-56126 Pisa, Italy}

\begin{document}
\title{Complexity of quantum motion
and quantum-classical correspondence: 
\\A phase-space approach}

\author{Jiaozi Wang}
\email{wangjz@mail.ustc.edu.cn}
\affiliation{\hefei}
\author{Giuliano Benenti}
\email{giuliano.benenti@uninsubria.it}
\affiliation{\como}
\affiliation{\infn}
\affiliation{\NEST}
\author{Giulio Casati}
\email{giulio.casati@uninsubria.it}
\affiliation{\como}
\affiliation{\brazil}
\author{Wen-ge Wang}
\email{wgwang@ustc.edu.cn}
\affiliation{\hefei}

\date{\today}

\begin{abstract}
We discuss the connection between the out-of-time-ordered correlator and the 
number of harmonics of the phase-space Wigner distribution function. 
In particular, we show that both 
quantities grow exponentially for chaotic dynamics, with a rate determined by the
largest Lyapunov exponent of the underlying classical dynamics, 
and algebraically - linearly or quadratically - for integrable dynamics. It is then possible to use such quantities 
to detect in the time domain the integrability to chaos crossover in many-body quantum systems. 
\end{abstract}

\maketitle

\emph{Introduction.-}
Understanding, characterizing, and measuring the complexity of (many-body) quantum 
dynamics is a fundamental problem, also of great practical relevance
for the prospects of quantum information science~\cite{qcbook} and more 
generally of quantum technologies.
In particular, the 
out-of-time-ordered correlator (OTOC) (see, e.g., \cite{Larkin96,Kitaev14,Maldacena14,Maldacena16,Hosur16,Galitski17,Prosen17,Fan17,Garttner17,Li17,Cotler18,Lin18,Jalabert18,Fazio18,Wei18,Dhar18,Richter18,Saraceno18,Sondhi18,Nahum18,Khemani18,Rakovszky18,Hirsch19,Cory18,Borgonovi19,Lakshminarayan19,Carlo19,Hirsch19b,Jalabert19}) has been put forward as a measure
of chaos in many-body quantum systems~\cite{Kitaev14}, and it has been related to mixing (called scrambling
when referring to quantum information) and thermalization, 
in different contexts, ranging from condensed matter~\cite{Larkin96} 
to black hole physics~\cite{Kitaev14,Maldacena14,Maldacena16}.

In classical mechanics a well defined notion of complexity exists, based on
local exponential instability of trajectories, that is, on the positivity of 
the largest Lyapunov exponent. In turn this implies positive algorithmic 
complexity~\cite{ford,alekseev}, 
so that orbits are in practice unpredictable, and 
memory of initial conditions is lost~\cite{lichtenberg}. 

The above notion of complexity cannot be readily transferred to quantum 
mechanics, where trajectories cannot be defined due to the Heisenberg 
uncertainty principle. On the other hand, complexity can be treated on equal
footing for classical and quantum mechanics in phase space,
see~\cite{Chirikov81,Gu90,Ford91,Gu97,Brumer97,Brumer03,harmonics08,harmonics09,harmonics10,prosen11,benenti12,harmonics14} for first attempts in this direction. 
The exponential sensitivity to initial conditions implies that the density distribution 
in the phase space is exponentially stretched and folded, and therefore 
becomes complex and intricate on ever smaller and smaller scales. 
Therefore, if one wants to reconstruct 
numerically the increasingly finer details of 
the phase-space distribution,  
it is intuitive to expect that the number of harmonics (i.e., components 
in Fourier space) that are excited also increases 
exponentially in time~\cite{Gu90,Brumer97,Brumer03}.
In an integrable system this does not happen, as the instability 
is typically linear in time~\cite{Casati80}.
It follows that the growth rate of the number of harmonics of the phase-space distribution
can be used, similarly to the Lyapunov exponent, as a way 
to characterize classical chaos.

One might object that there is the problem of choosing the 
phase-space coordinates. However, while the finite-time results of course depend on 
the basis, the asymptotic growth rate does not, as it is given by the
largest Lyapunov exponent~\cite{footnote_generalized}, which is base-independent.   
If we consider action-angle variables of an unperturbed, integrable system, 
then a non-integrable perturbation will lead to an exponential growth of the number of harmonics
with a rate given by the largest Lyapunov exponent. On the other hand, 
an integrable perturbation may distort even strongly unperturbed tori, so that at short times a large 
number of harmonics may be excited, but asymptotically the number of harmonics will grow 
linearly in time. 

The phase-space approach can be conveniently generalized to quantum mechanics,
using the number of harmonics of the Wigner function as a suitable measure
of the complexity of a quantum state. 

In this Letter we study the number of harmonics and the OTOC, 
both for classically chaotic and integrable systems. 
We show that both quantities can be used to detect, in the time domain, the 
crossover from integrability to chaos. 
Moreover, we show that the number of harmonics and the OTOC are connected.
Such connection places the OTOC on a broader context and reinforces its 
interpretation as a complexity quantifier.
The above results are illustrated numerically for the case of 
two nonlinearly coupled oscillators.

\emph{Theoretical basis.-}
We start by considering the number of harmonics in classical mechanics.
Let $H=H_0+H_I$ be the Hamiltonian of a $N$-particle system,
where $H_0$ is time-independent and integrable, 
while $H$ might be either integrable or non-integrable.
We write the classical distribution function $\rho(\boldsymbol{I},\boldsymbol{\theta};t)$
in terms of the action-angle variables of $H_0$, $\boldsymbol{I}=(I_1,...,I_N)$ and 
$\boldsymbol{\theta}=(\theta_1,...,\theta_N)$. 
Taking the Fourier transform 
\begin{equation}\label{eq-rhoc-ft}
\rho(\boldsymbol{I},\boldsymbol{\theta};t)=\frac{1}{\pi^{N}}\sum_{\boldsymbol{m}}\rho_{\boldsymbol{m}}(\boldsymbol{I};t)\exp(i\boldsymbol{m\cdot\theta}),
\end{equation}
we define the second moment of the harmonics distribution as 
\be\label{eq-M2cl-I}
{\cal M}_{2}^{cl}(t)
=\frac{\sum_{\boldsymbol{m}} \boldsymbol{m}^{2}\int_{0}^{\infty}dI|\rho_{\boldsymbol{m}}(\boldsymbol{I};t)|^{2}}{\sum_{\boldsymbol{m}}\int_{0}^{\infty}dI|\rho_{\boldsymbol{m}}(\boldsymbol{I};t)|^{2}}.
\ee
Making use of the properties of the Fourier transform, 
${\cal M}^{cl}_2 (t)$ can also be written as
\be\label{eq-m2cl}
{\cal M}_{2}^{cl}(t)=\frac{\int d\boldsymbol{I}d\boldsymbol{\theta}\sum_{k}\left|\frac{\partial}{\partial\theta_{k}}\rho(\boldsymbol{I},\boldsymbol{\theta};t)\right|^{2}}{\int d\boldsymbol{I}d\boldsymbol{\theta}\left|\rho(\boldsymbol{I},\boldsymbol{\theta};t)\right|^{2}}.
\ee
Due to the Hamiltonian evolution, the denominator of Eq.(\ref{eq-m2cl}) is constant in time.
So the behavior of ${\cal M}_2^{cl}(t)$ is determined by the  numerator, notably by the behavior of $\frac{\partial}{\partial\theta_{k}}\rho(\boldsymbol{I},\boldsymbol{\theta};t)$ for different $k$. We obtain~\cite{inprep}
\be\label{eq-M2cl-3}
{\cal M}_{2}^{cl}(t)=\frac{\int d\boldsymbol{\chi}(0)\left|\widetilde{\nabla}\rho_{0}(\boldsymbol{\chi}(0))\right|^{2}\sum_{k=1}^{N}(\delta I_{k}(t))^{2}}{d^{2}\int d\boldsymbol{\chi}(0)\left|\rho_0(\boldsymbol{\chi}(0))\right|^{2}},
\ee
where $\boldsymbol{\chi}(t)=\ensuremath{(\theta_{1}(t),\cdots,\theta_{N}(t),I_{1}(t),\cdots,I_{N}(t))}$ is the action-angle vector at time $t$, 
 $\widetilde{\boldsymbol{\nabla}}=(\frac{\partial}{\partial I_{1}},\cdots,\frac{\partial}{\partial I_{N}},-\frac{\partial}{\partial\theta_{1}},\cdots,-\frac{\partial}{\partial\theta_{N}})$ a modified gradient, $\rho_{t}(\boldsymbol{\chi}(t))$
the classical distribution function,
and $\delta I_{k}(t)$ indicates the deviation of the $k$-th action at time $t$, for two points in phase space, whose  initial positions are $\boldsymbol{\chi}(0)$ and $\boldsymbol{\chi}(0)+\delta\boldsymbol{\chi}(0)$, respectively, with $\delta\boldsymbol{\chi}(0)=d\frac{\boldsymbol{\widetilde{\nabla}}\rho_{0}(\boldsymbol{\chi}(0))}{|\boldsymbol{\widetilde{\nabla}}\rho_{0}(\boldsymbol{\chi}(0))|}$ and $d\to 0$.
 From Eq.~(\ref{eq-M2cl-3}) one finds that the behavior of the number of harmonics in classical case is determined by the quantities $\delta I_{k}(t)$.
In the integrable case, $\delta I_{k}(t)$ typically increases linearly with time,
and thus ${\cal M}_2 ^{cl}(t)$ has a quadratic growth.
For the chaotic case,  $\delta I_{k}(t)$ increase exponentially with time, 
and so is the case for ${\cal M}_{2}^{cl}(t)$. 
For a numerical confirming evidence, see below.

In the quantum case, for the sake of simplicity we limit ourselves to
systems whose Hamiltonian $\hat{H}$ can be written in terms of a set
of bosonic creation-annihilation operators. That is, 
$\hat{H}= \hat{H}_{0}+\hat{H}_{1}$,
 where $\hat{H}_{0}=\hat{H}_{0}(\hat{n}_{1},\cdots,\hat{n}_{N})$
is time-independent and integrable, while
 $\hat{H}_{1}=\hat{H}_{1}({\hat{a}_{1}^{\dag},
\cdots,\hat{a}_{N}^{\dag}},\hat{a}_{1},\cdots,
 \hat{a}_{N};t)$, with 
$[\hat{a}_{i},\hat{a}_{j}]=[\hat{a}_{i}^{\dag},\hat{a}_{j}^{\dag}]=0$,
$[\hat{a}_{i}^{\dag},\hat{a}_{j}]=\delta_{ij}$, and
the number operators $\hat{n}_i= \hat{a}_i^\dag \hat{a}_i$.
The Wigner function of a state, which is described
 by a density operator $\hat{\rho}(t)$, can be written 
as~\cite{Bargmann,Glauber,Agarwal}
$$
W({\boldsymbol\alpha},{\boldsymbol\alpha}^{*};t)=\frac{1}{\pi^{2N}
\hbar^N} \int d^{2}{\boldsymbol\eta}
\exp{\left(\frac{{\boldsymbol\eta}^*\cdot{\boldsymbol\alpha}}{\sqrt{\hbar}}
-\frac{{\boldsymbol\eta}
\cdot{\boldsymbol\alpha}^*}{\sqrt{\hbar}}\right)}
$$
\begin{equation}
\times \text{Tr}[
\hat{\rho}(t)\hat{D}({\boldsymbol\eta})],\label{wignerm}
\end{equation}
where ${\boldsymbol\eta}=(\eta_1,...,\eta_N)$ and
${\boldsymbol\alpha}=(\alpha_1,...,\alpha_N)$ are $N$-dimensional
complex variables, and the displacement operator 
\begin{equation}
\hat{D}\left({\boldsymbol\eta}\right)=
\exp\left[\sum_{i=1}^{N}{\left(\eta_i\hat{a_{i}}^
{\dagger}-\eta_i^{*}\hat{a_{i}}\right)}\right].
\end{equation}

We then consider the Fourier expansion of the Wigner function,
\be\label{eq-fourierW}
W(\boldsymbol{\alpha},\boldsymbol{\alpha}^{*};t)=\frac{1}{\pi^{N}}\sum_{\boldsymbol{m}}W_{\boldsymbol{m}}(\boldsymbol{I};t)\exp(i\boldsymbol{m}\cdot\boldsymbol{\theta}).
\ee
Here $I_k$ and $\theta_k$ are action and angle variables of the unperturbed Hamiltonian
$H_0$, 
with $\alpha_{k}=\sqrt{I_{k}}\exp(i\theta_{k})$.
The number of harmonics is estimated by 
$\sqrt{{\cal M}_2}$, where 
${\cal M}_2 (t)=\sum_{\boldsymbol{m}}|\boldsymbol{m}|^{2}{\cal W}_{\boldsymbol{m}}(t)$ is the second moment of the harmonics distribution, with 
\be\label{eq-rhoeta}
{\cal W}_{\boldsymbol{m}}(t)=\frac{\int d\boldsymbol{I}|W_{\boldsymbol{m}}(\boldsymbol{I};t)|^{2}}{\sum_{\boldsymbol{m}}\int d\boldsymbol{I}|W_{\boldsymbol{m}}(\boldsymbol{I};t)|^{2}}.
\ee
In terms of density matrix $\hat{\rho}(t)$, we can also write~\cite{Gu90,Brumer97,harmonics08,harmonics10} 
\be\label{eq-Wm}
{\cal W}_{\boldsymbol{m}}(t)=\frac{\sum\limits _{n_{k}>0}|\langle\boldsymbol{n}+\boldsymbol{m}|\hat{\rho}(t)|\boldsymbol{n}\rangle|^{2}}{\sum\limits _{n_{k}>0,}\sum\limits _{(m_{k}+n_{k})>0}|\langle\boldsymbol{n}+\boldsymbol{m}|\hat{\rho}(t)|\boldsymbol{n}\rangle|^{2}}.
\ee
Following~\cite{Gu90,Brumer03},
${\cal M}_2(t)$ can be written as
\be\label{eq-rhoI}
{\cal M}_{2}(t)=\frac{\sum_{k}\text{Tr}(|[\hat{\rho}(t),\hat{I}_{k}]|^{2})}{\hbar^{2}\text{Tr}(\hat{\rho}^{2}(t))},
\ee 
where $\hat{I}_k=\hbar \hat{n}_k$. In 
the case of a pure state $|\psi(t)\rangle$,
\be
{\cal M}_{2}(t)=\frac{2\sum_{k}(\Delta I_k(t))^2}{\hbar^2},
\ee
where
\be
(\Delta I_k(t))^2=\langle\psi(t)|\hat{I}_{k}^{2}|\psi(t)\rangle-\langle\psi(t)|\hat{I}_{k}|\psi(t)\rangle^{2}
\ee
is the variance for measurement at time $t$, corresponding to observable $\hat{I}_k$.

We now turn to OTOC, defined as the expectation of the square commutator of two operators
taken at different times:
\be\label{eq-OTOC-form}
{\cal C}(t)=\left\langle |[\hat{A}(t),\hat{B}(0)]|^{2}\right\rangle.
\ee
It is interesting to remark that the numerator of the rhs of 
Eq.(\ref{eq-rhoI}), which measures the number of harmonics (the denominator is constant in time for Hamiltonian evolution), turns out to be a particular kind of OTOC,
with $\hat{A}(t)= \hat{\rho}(t)$ and $\hat{B}(0)= \hat{I_k}$, 
and the average taken over the whole Hilbert space.
In the following however, we consider a more commonly used expression for OTOC,
\begin{equation}
C_{pp}(t)=-\frac{1}{\hbar^2}\langle\psi_{0}|[\hat{p}_{1}(t),\hat{p}_{1}(0)]^{2}|\psi_{0}\rangle,
\end{equation}
where $\hat{p}_1$ is the momentum operator for one particle 
(denoted as particle 1) and 
$|\psi_{0}\rangle$ is the initial state (at time $t=0$) of the composite,
$N$-particle system.

The  classical correspondence of $C_{pp}(t)$, denoted as $C_{pp}^{cl}(t)$,
is obtained by the canonical substitution
$\frac{1}{i\hbar}[\hat{A},\hat{B}]\rightarrow \{A,B\}_{\text{PB}}$,
where the right-hand side is the Poisson bracket of the classical 
variables $A$ and $B$.
We obtain
\begin{align}\label{eq-OTOCCL}
C_{pp}^{cl}(t) & =\int 
d\boldsymbol{\gamma}(0)\rho_{0}(\boldsymbol{\gamma}(0))
\{p_{1}(t),p_{1}(0)\}_{PB}^{2} \nonumber \\
& =\int d\boldsymbol{\gamma}(0)\rho_{0}(\boldsymbol{\gamma}(0))\left(\frac{\delta p_{1}(t)}{\delta q_{1}(0)}\right)^{2},
\end{align}
where $\boldsymbol{\gamma}(t)=(q_{1}(t),\cdots,q_{N}(t),p_{1}(t),\cdots,p_{N}(t))$ and
$\rho_{0}(\boldsymbol{\gamma}(0))$ is the 
initial distribution of the classical ensemble corresponding,
in the quantum case, to the Wigner function of the initial state.
It is clear from Eq.(\ref{eq-OTOCCL}) that the behavior of 
$C_{pp}^{cl}(t)$ is determined by the quantity 
$\left(\frac{\delta p_{1}(t)}{\delta q_{1}(0)}\right)^{2}$.
In the integrable case, $\delta p_1(t)$ increases linearly with time,
and therefore $C_{pp}^{cl}(t)$ is expected to be a quadratic function 
of time, while in the chaotic case  
${\delta p_{1}(t)}$ has an exponential growth, 
and so is $C_{pp}^{cl}(t)$.

Barring subtle points related to the phase-space averaging procedure~\cite{inprep}, 
we can conclude that both ${\cal M}_{2}^{cl}(t)$ 
and the OTOC $C_{pp}^{cl}(t)$ grow exponentially for
chaotic classical systems, with a rate given by twice the 
largest Lyapunov exponent
$\lambda_L(\boldsymbol{r}_{0})$ of the trajectory started at the 
center of the initial distribution, assumed to be narrow:
\be\label{eq-COM-OTOC}
{\cal M}_{2}^{cl}(t),C_{pp}^{cl}(t)
\propto\exp(2\lambda_{L}(\boldsymbol{r}_{0})\,t).
\ee 
While the short-time behavior depends on the chosen phase-space coordinates,
the growth rate, which is determined by the largest Lyapunov exponent, does not.

\emph{Numerical results.-}
To illustrate the above analytical results, we consider a 
model of two coupled oscillators, whose Hamiltonian is 
\begin{equation}
H=\frac{1}{2}(\hat{p}_{1}^{2}+\hat{p}_{2}^{2})+
\frac{\beta}{4}(\hat{q}_{1}^{4}+\hat{q}_{2}^{4})+
\frac{1}{2}\hat{q}_{1}^{2}\hat{q}_{2}^{2}.
\end{equation}
This model allows us to investigate both the 
integrable case $\beta=1$, the chaotic case $\beta\ll 1$, and the crossover 
between the two regimes.
As initial state $|\psi_0\rangle$ we consider the tensor product of Gaussian 
wave packets for both particles. 
We average our numerical data over $\mathcal{N}$ initial conditions, 
with the centers of the (Gaussian) Wigner functions randomly distributed 
within the energy shell $[E_0-\delta E/2,E_0+\delta E/2]$, with $\delta E\ll E$. In our dimensionless units the width in $q_k$ and $p_k$ ($k=1,2$) of the
initial Wigner functions is proportional to 
$\sqrt{\hbar}$, with $\hbar$ effective Planck constant. 
In the classical case, we average over the corresponding initial Gaussian
distributions~\cite{num_method}.
 
\begin{figure}
        \includegraphics[width=1\linewidth]{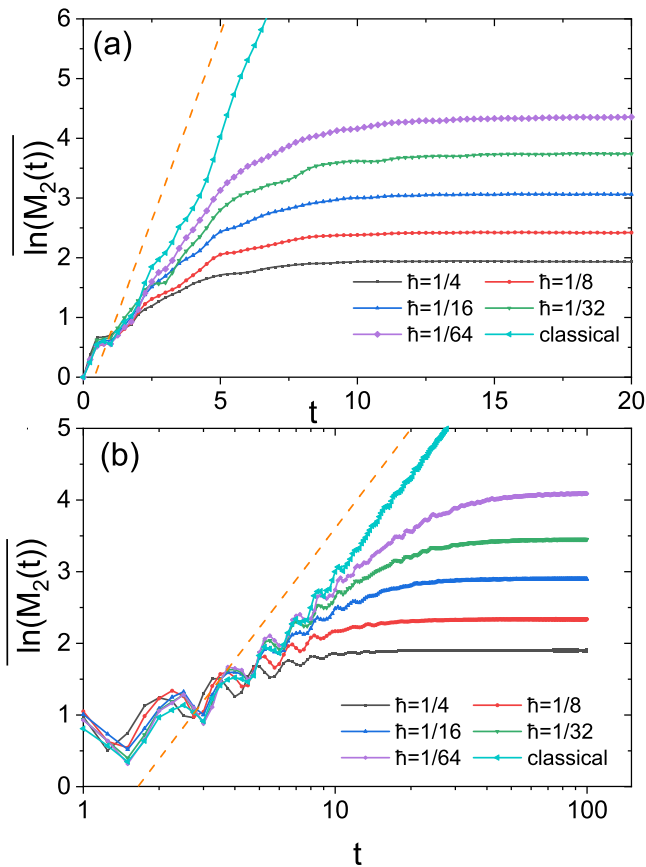}
        \caption{Dependence of the averaged number of harmonics on time 
for different value of $\hbar$ and in the classical case
(a) in the chaotic regime for $\beta=0.1$  
(the dashed line 
indicates the exponential growth of 
${\cal M}_2$, with rate $1.25$ equal to twice a generalized 
largest Lyapunov exponent); (b) in the integrable regime $\beta=1$, 
where $t$ is given in logarithmic scale (the dashed line 
indicates the quadratic growth of ${\cal M}_2$). In the classical case we average over 100 different initial ensembles and in the quantum case over 100 initial states, distributed randomly within the energy shell with $E_0=5,\delta E=0.002$. To avoid effects due to the size of $\hbar$, which are not relevant 
to determine the growth rate, here 
we subtract the $t=0$ value.}
\label{M2_compare}
 \end{figure}

\begin{figure}
	\includegraphics[width=1\linewidth]{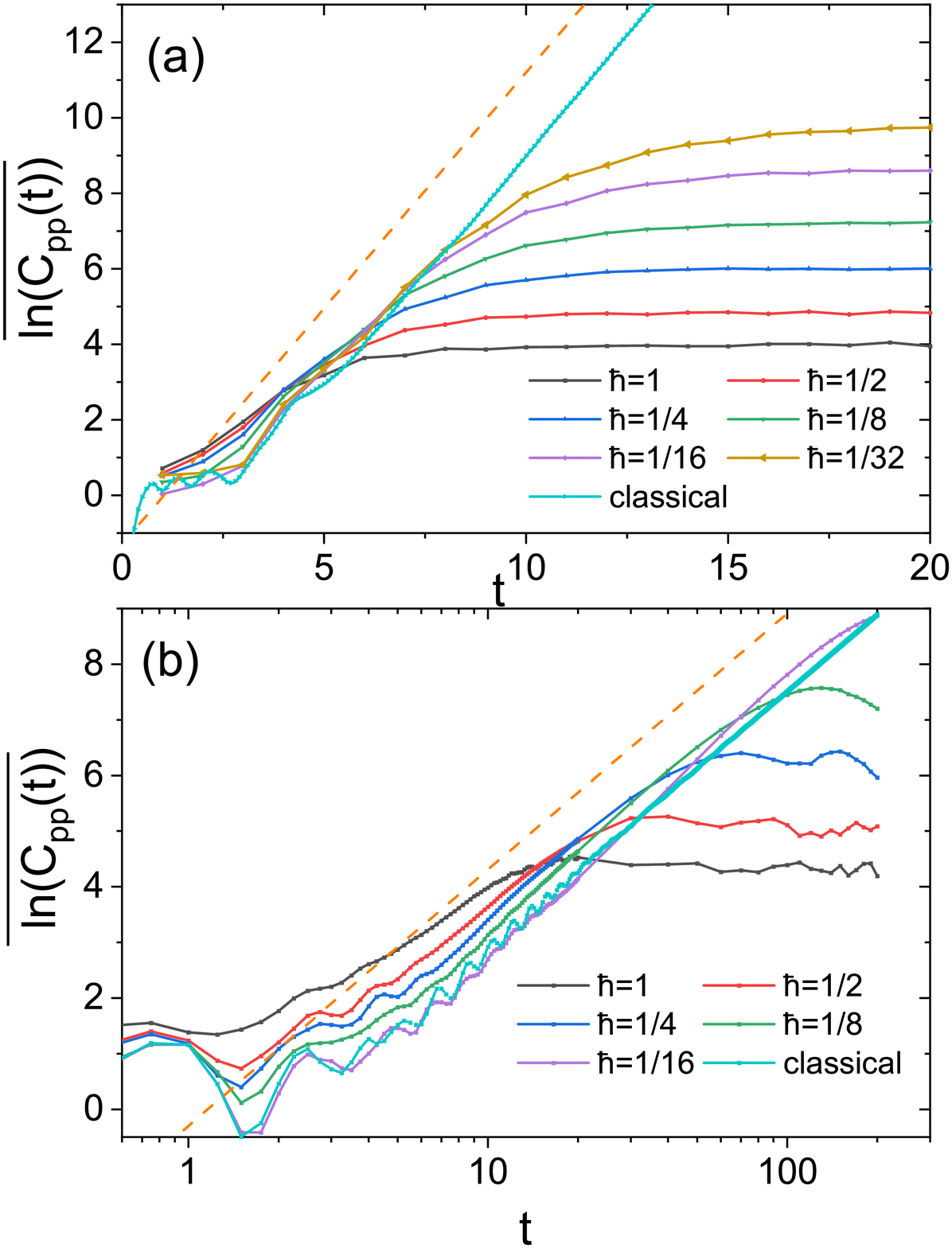}
	\caption{Dependence of averaged OTOC on time for different value of $\hbar$ and in
the classical case, (a) in the chaotic regime for $\beta=0.1$;
(b) in the integrable regime $\beta=1$.
In the classical case, the results are averaged over 100 different initial ensemble and in quantum case over $100$ initial states, respectively, distributed randomly in the same energy shell as in the previous figure.
The dashed lines indicate exponential (top) and quadratic (bottom) growth.
In the chaotic case, the growth rate is the same as the one in 
Fig.~\ref{M2_compare} for the number of harmonics.}
\label{OTOC_All}
\end{figure}

Numerical results for the number of harmonics and the OTOC are shown 
in Fig.~\ref{M2_compare} and Fig.~\ref{OTOC_All}, respectively.
Here we average over different initial states as follows: 
\begin{align}
\overline{\ln{{\cal M}_2(t)}} & =\frac{1}{\mathcal{N}}\sum_{k=1}^{\mathcal{N}} 
\ln{{\cal M}_2^{(k)}(t)}, \nonumber \\
\overline{\ln{C_{pp}(t)}} & =\frac{1}{\mathcal{N}}
\sum_{k=1}^{\mathcal{N}} \ln{C_{pp}^{(k)}(t)},
\label{eq:averaging}
\end{align}
where ${\cal M}_2^{(k)}(t)$ and $C_{pp}^{(k)}(t)$ 
are the number of harmonics and OTOC starting from the $k$-th 
initial condition, and $\mathcal{N}$, as discussed above,
is the number of initial conditions we consider within the energy shell.
Such averaging addresses the problem of spurious exponential growth  
in integrable systems, due to unstable fixed points in phase 
space rather than to chaos~\cite{otoc-Cao20,Hirsch19b}. 
With the averaging (\ref{eq:averaging}),
we can unambiguously associate the 
exponential (quadratic) growth of ${\cal M}_2(t)$ and $C_{pp}(t)$ 
to chaos (integrability).
Indeed, in agreement with theoretical predictions, 
it is seen that both quantities, in the semiclassical regime and 
for chaotic dynamics, 
grow exponentially, with a rate 
determined by the largest Lyapunov exponent 
$\lambda_L$~\cite{footnote_generalized,inprep}. 
On the other hand, the growth is 
quadratic when the system is integrable.
In all cases, it is clear that the quantum results converge to the 
classical ones when the effective Planck constant $\hbar\to 0$. 
For the OTOC, it appears that the quantum to classical correspondence, as expected, 
is valid up to the logarithmic time scale in the chaotic case 
$t_E\sim \ln(1/\hbar)/\lambda_L$, while in 
the integrable case the correspondence is valid up to a much longer
time, growing as $1/\hbar$. 
With regard to the number of harmonics, it can be seen
that the convergence of quantum results to the classical ones 
requires smaller values of $\hbar$.
The deviation of ${\cal M}_{2}(t)$ from ${\cal M}^{cl}_{2}(t)$ 
is determined by the deviation of the Wigner function from the corresponding 
classical distribution function. Though we consider initial coherent states, 
for which the Wigner function is everywhere positive and exactly equal 
to the corresponding classical distribution, nonlinear dynamics 
generates negative fringes in the Wigner function, 
which usually deviates from the classical distribution function at times 
much shorter than the Ehrenfest time.
Thus to reveal the integrable or chaotic nature of dynamics from 
the quadratic or exponentially growth of the number of harmonics,
one needs a smaller $\hbar$ than for the OTOC.

\begin{figure}
	\includegraphics[width=1\linewidth]{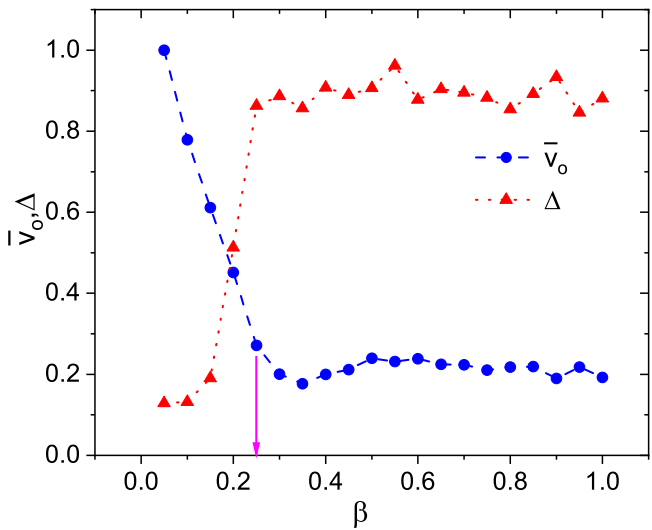}
	\caption{Average velocity $\overline{v}_o$  in the growth of OTOC and distance to chaos measured by $\Delta$.
Data for $\Delta$ are for $\hbar=1/32$, while for 
$\overline{v}_o$ $\hbar=1/8$.
To get a better comparison of the results, $\overline{v}_o$ is rescaled to its maximal value. The arrow roughly indicates the value of $\beta$ for the transition to chaos.}
\label{DV}
\end{figure}

The different time dependence between integrable and chaotic regime 
for number of harmonics and OTOC suggests that these quantities can be used to detect
the crossover from integrability to chaos in the time domain.
In order to provide a clear evidence of this expectation, we plot in Fig.~\ref{DV} an average velocity
$\overline{v}_o=\overline{C_{pp}}(t_o^\star)/t_o^\star$, 
where $t_o^\star$ indicates the time at which $\overline{C_{pp}}$ reaches half of 
its saturation value. The figure shows that 
 the crossover
to chaos takes place in the region around $\beta=0.25$. 
This is in agreement with results obtained from  
energy-level statistics~\cite{haake}, which is the ordinary tool to detect such a transition. 
In the latter case, we consider a parameter $\Delta$, defined as
\be
\Delta=\frac{\int_{0}^{\infty}|P(s)-P_{w}(s)|ds}{\int_{0}^{\infty}|P_{p}(s)-P_{w}(s)|ds},
\ee
where $P(s)$ is the nearest-level-spacing distribution of the system, and 
$P_w(s)$($P_p(s)$) is the Wigner-Dyson (Poisson) distribution. 
The quantity $\Delta$ measures the distance 
between the level spacings distribution of our model and the 
Poisson and Wigner-Dyson distributions, normalized in such a way that $\Delta=0$ 
corresponds to Wigner-Dyson and $\Delta=1$ to 
Poisson distribution. Variation of $\Delta$ versus $\beta$ shows the same crossover region around $\beta=0.25$.
 
\emph{Discussion and conclusions.-}
We have shown that both number of harmonics and OTOC are convenient tools to characterize
complexity in classical and quantum mechanics. Both quantities can be used, similarly to 
spectral statistics, to detect the crossover from integrability to chaos.  
In contrast with spectral statistics, which can only provide information on the integrable or
chaotic nature of dynamics, OTOC and number of harmonics also measure the ``strength'' of chaos,
as provided by the largest Lyapunov exponent. 
In this connection, we recall that non-integrable systems with zero Lyapounov exponent (e.g. irrational polygonal billiards) exhibit Wigner-Dyson statistics. 

While in classical mechanics the (exponential or quadratic) growth of both quantities 
continues forever, in quantum mechanics it eventually saturates due to the 
discreteness of quantum phase space, i.e., its non-commutative geometry.
In particular, for chaotic dynamics and in the semiclassical regime the saturation value
of number of harmonics and OTOC can be obtained  
using the eigenstate thermalization hypothesis~\cite{Sred99}.
We obtain~\cite{inprep} 
$\overline{\ln{\cal M}_{2}(\infty)},\overline{\ln C_{pp}(\infty)}\propto-\ln\hbar$.
Note that, as the growth of number of harmonics and OTOC is exponential in time, 
saturation takes place on the time scale logarithmically
short in $\hbar$. 

It is also interesting to comment on the relationship between OTOC and number of harmonics. 
While in the quantum case the link can be understood by interpreting the second moment
of the harmonics distribution as a particular kind of OTOC, it is the phase space picture 
which allows a more intuitive connection. When considering canonical variables, the classical 
OTOC is the phase-space average of their squared Poisson bracket, which in turn is related to exponential sensitivity 
of trajectories, 
see Eq.~(\ref{eq-OTOCCL}) for the case of $C_{pp}$. 
On the other hand, the number of harmonics can be seen as the phase-space average of 
$\sum_k (\delta I_k(t))^2$, with $I_k$ action variables for the unperturbed Hamiltonian,  
over a modified distribution function, see Eq.~(\ref{eq-M2cl-3}).
For both quantities, it is the integrable or non-integrable nature 
of the perturbation which determines their growth in time. 
While the relationship between OTOC and number of harmonics is generic, an intuitive understanding of their connection on the basis of the phase-space approach still has to be developed for many-body quantum systems without classical analogue, i.e. for quantum spin chains.

The last point we want to mention here is that the second moment 
${\cal M}_{2}(t)$ of the 
harmonics distribution 
can in principle be mesured experimentally, by making 
used of  the multiple quantum coherence (MQC) experiments, which is a method 
already used to measure OTOC~\cite{MQC}.
To measure ${\cal M}_{2}(t)$, one can adapt the method of 
Ref.~\cite{MQC_Rey}. To simplify writing, we consider the 
one-dimensional case.    
Let $|n\rangle$ be the eigenstates of the number operator $\hat{n}$: 
$\hat{n}|n\rangle=n|n\rangle$.
One can divide the density matrix of an arbitrary state $\hat{\rho}(t)$
into blocks as $\hat{\rho}(t)=\sum_{m}\hat{\rho}_{m}(t)$,
where 
\be
\hat{\rho}_{m}(t)=\sum_{n-n^{\prime}=m}\rho_{nn^{\prime}}(t)|n\rangle\langle n^{\prime}|.
\ee
The Frobenius
norm $I_{m}(\hat{\rho}(t))=(\left\Vert \hat{\rho}_{m}(t)\right\Vert )^{2}=\text{tr}[\hat{\rho}_{m}^{\dagger}(t)\hat{\rho}_{m}(t)]=\sum_{n-n^{\prime}=m}|\rho_{nn^{\prime}}(t)|^{2}$, is related to ${\cal W}_{m}(t)$ (see Eq.~(\ref{eq-Wm})) 
as follows:
\be
I_{m}(\hat{\rho}(t))=\frac{{\cal W}_{m}(t)}{\text{tr}(\hat{\rho}^{2}(t))}.
\ee
To measure $I_{m}(\hat{\rho}(t))$, one can evovle
$\hat{\rho}_{0}$ into $\hat{\rho}(t)$, apply 
$\hat{W}(\phi)=e^{-i\hat{n}\phi}$, evovle
backward (echo experiment) to $\hat{\rho}_{f}$, and measure
the probability to find the system in the initial state,
given by $\text{tr}[\hat{\rho}_{0}\hat{\rho}_{f}]$.
Since $\hat{W}(\phi)\hat{\rho}_{m}(t)\hat{W}^{\dagger}(\phi)=
e^{im\phi}\hat{\rho}_{m}(t)$, we obtain
\be
\text{tr}[\hat{\rho}_{0}\hat{\rho}_{f}]=
\sum_{m}I_{m}(\hat{\rho}({t}))e^{-im\phi}.
\ee
Finally, by Fourier transforming the signal with respect to $\phi$,
one obtains the MQC spectrum $\{I_{m}(\hat{\rho}({t}))\}$, and 
the second moment of the harmonics distribution:
\be
{\cal M}_{2}(t)={\text{tr}}(\hat{\rho}_{m}^{2}(t))\sum_{m}m^{2}I_{m}
(\hat{\rho}(t)).
\ee

{\it Acknowledgments}:
J.W. and W-G.W. acknowledge the Natural Science Foundation of China under Grant
Nos.~11535011, and 11775210.
G.B. acknowledges the financial support of the INFN through the project “QUANTUM”.


\begin{thebibliography}{99}
\bibitem{footnote_generalized}
More precisely, one should consider a generalized Lyapunov exponent, see \cite{Brumer97,Brumer03}.
Moreover, we do not consider here patological cases like the baker's map for which 
orbits are exponentially unstable but Liouville phase-space evolution is exactly 
solvable~\cite{prosen11}.
\end{thebibliography}
\end{document}